\documentclass[twocolumn,floatfix,aps,superscriptaddress]{revtex4-2}
\usepackage{graphicx}
\usepackage{amsmath}
\usepackage{amssymb}
\usepackage{bm}
\usepackage{hyperref}

\usepackage{color}

\begin{document}

\title{Majorana-metal transition in a disordered superconductor:\\
percolation in a landscape of topological domain walls}
\author{V. A. Zakharov}
\email{zakharov@lorentz.leidenuniv.nl}
\affiliation{Instituut-Lorentz, Universiteit Leiden, P.O. Box 9506, 2300 RA Leiden, The Netherlands}
\author{I. C. Fulga}
\affiliation{Institute for Theoretical Solid State Physics, IFW Dresden, Germany}
\affiliation{W\"{u}rzburg-Dresden Cluster of Excellence ct.qmat, Germany}
\author{G. Lemut}
\affiliation{Dahlem Center for Complex Quantum Systems and Physics Department,
Freie Universit\"{a}t Berlin, Arnimallee 14, 14195 Berlin, Germany}
\author{J. Tworzyd{\l}o}
\affiliation{Faculty of Physics, University of Warsaw, ul.\ Pasteura 5, 02--093 Warszawa, Poland}
\author{C. W. J. Beenakker}
\affiliation{Instituut-Lorentz, Universiteit Leiden, P.O. Box 9506, 2300 RA Leiden, The Netherlands}

\date{October 2024}

\begin{abstract}
Most superconductors are thermal insulators. A disordered chiral \textit{p}-wave superconductor, however, can make a transition to a thermal metal phase. Because heat is then transported by Majorana fermions, this phase is referred to as a Majorana metal. Here we present numerical evidence that the mechanism for the phase transition with increasing electrostatic disorder is the percolation of boundaries separating domains of different Chern number. We construct the network of domain walls using the spectral localizer as a ``topological landscape function'', and obtain the thermal metal--insulator phase diagram from the percolation transition.
\end{abstract}
\maketitle

\section{Introduction}

While a superconductor is a perfect conductor of electricity, it generally conducts heat poorly. Adding disorder is not expected to improve this, but in a  two-dimensional (2D) superconductor with chiral \textit{p}-wave pairing \cite{Kal15} the unexpected happens: If sufficiently many defects are added the thermal insulator becomes a thermal metal \cite{Sen00,Eve08,Bee16}. This unusual state is known as a Majorana metal, because the quasiparticles that conduct the heat are Majorana fermions (equal-weight superpositions of electrons and holes). Although the transition from a thermal insulator to a thermal metal has not yet been observed in experiments, it has been demonstrated in computer simulations \cite{Cha02,Mil07,Wim10,Kag10,Med11,Lau12,Pek19,Wan21}.

The Majorana-metal transition is well understood if the defects consist of the Abrikosov vortices that appear when a perpendicular magnetic field is applied to a type-II superconductor. A vortex can bind sub-gap quasiparticles \cite{Car64}, but bound states in nearby vortices will not typically be aligned in energy, making them inefficient for heat transport. A special property of a chiral \textit{p}-wave superconductor is that its vortices have a bound state exactly in the middle of the gap ($E=0$, the Fermi level), a socalled Majorana zero-mode \cite{Vol99,Rea00,Bee13,Lut18,Fle21}. The energetic alignment of Majorana zero-modes allows for resonant heat conduction when the density of Abrikosov vortices crosses a critical threshold \cite{Cha02,Mil07}.

Electrostatic disorder in zero magnetic field can also produce a thermal metal phase \cite{Wim10}. The phase transition falls in the same universality class D as for vortex disorder \cite{Wan21}, and one would expect the mechanism to be related in the same way to the appearance of Majorana zero-modes --- even without any vortices to bind them. Can we demonstrate that in a computer simulation?

To address this question we use the spectral localizer approach pioneered by Loring and Schulz-Baldes \cite{Lor15,Ful16,Lor17,Loz19,Lor20,Cer22,Sch24,Dol24}. The spectral localizer embeds the Hamiltonian $\pm H$ on the diagonal of a $2\times 2$ matrix, with the position operator $x\pm iy$ on the off-diagonal. Its spectrum quantifies whether Hamiltonian and position can be made commuting by a deformation that does not close the excitation gap \cite{Fra24,Qi24}.

In a class D system the matrix signature of the spectral localizer (number of positive minus number of negative eigenvalues) identifies domains of different Chern number \cite{Loz19,Dol24}. As discussed by Volovik \cite{Vol18}, the domain walls support low-lying states at energy $E\simeq \hbar v_{\rm F}/\ell$ for a domain of linear dimension $\ell$. These states become Majorana zero-modes in the limit $\ell\rightarrow \infty$ of a percolating domain wall. By identifying the metal-insulator transition with the percolation transition of the domain walls we construct the phase diagram in a closed system, and compare with calculations based on the thermal conduction in an open system \cite{Wim10,Wan21}.

\section{Topological landscape function}

\begin{figure*}[tb]
\centerline{\includegraphics[width=0.7\linewidth]{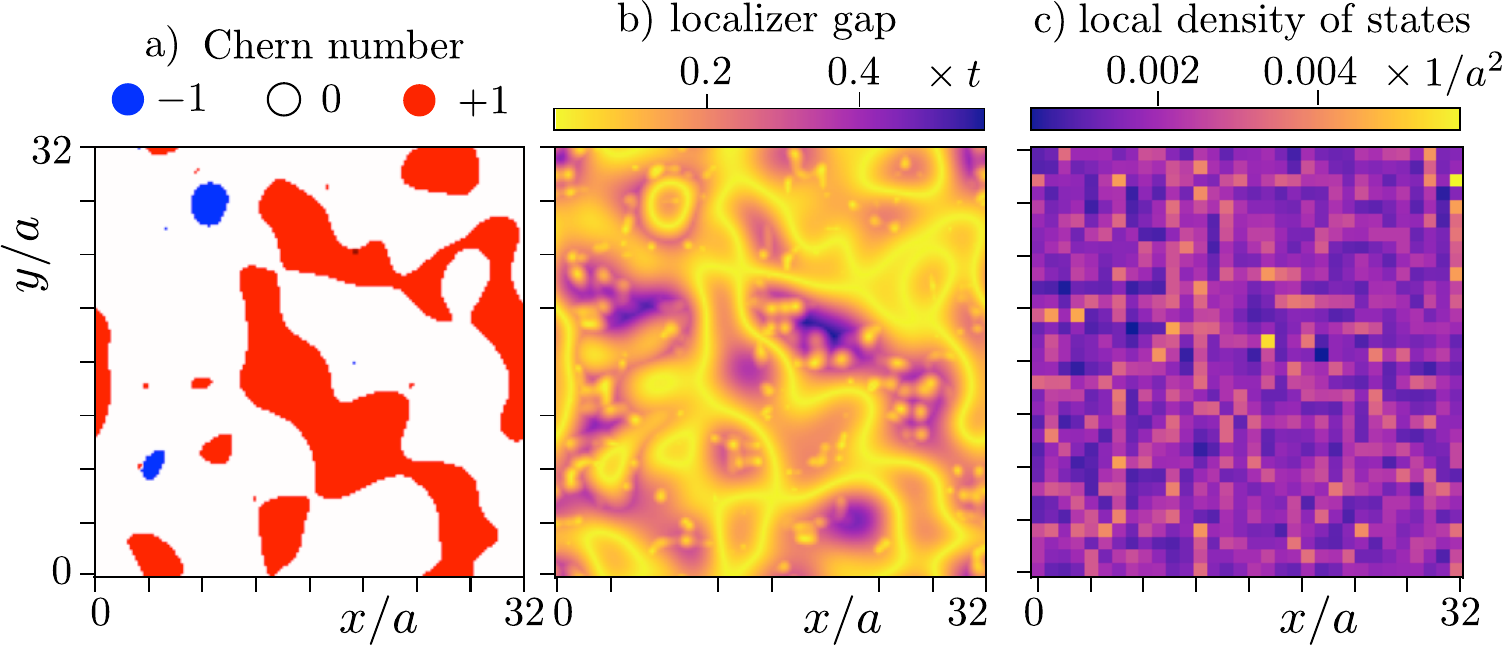}}
\caption{Panels a) and b) show the topological landscape function [Chern number ${\cal C}(x,y)$ and localizer gap $\delta(x,y)$] in a disordered chiral \textit{p}-wave superconductor (Hamiltonian \eqref{Hlattice}, parameters $\Delta=4t$, $\bar{\mu}=t$, $\delta\mu=4\,t$, $L=32a$, periodic boundary conditions). At these parameters the superconductor is in the thermal metal phase. Panel c) shows that the network of domain walls leaves no trace in the local density of states (integrated over the energy interval $|E|<0.2\,t$).
}
\label{fig_landscape}
\end{figure*}

\subsection{Lattice Hamiltonian}

The Bogoliubov-De Gennes Hamiltonian for a chiral \textit{p}-wave superconductor is
\begin{equation}
H_{\rm BdG}=\begin{pmatrix}
p^2/2m-E_{\rm F}&v_\Delta(p_x-ip_y)\\
v_\Delta(p_x+ip_y)&E_{\rm F}-p^2/2m
\end{pmatrix}.
\end{equation}
It acts on a two-component wave function $\Psi=(\psi_{\rm e},\psi_{\rm h})$, the pair potential $\propto v_\Delta$ couples the electron and hole components (filled states above the Fermi level $E_{\rm F}$, respectively, empty states below $E_{\rm F}$, with $E_{\rm F}=\tfrac{1}{2}mv_{\rm F}^2$ in terms of the effective mass $m$ and Fermi velocity $v_{\rm F}$). Because this is equal-spin pairing, we can omit the spin degree of freedom. 

The particle-hole symmetry relation,
\begin{equation}
\sigma_x H^\ast_{\rm BdG}\sigma_x=-H_{\rm BdG},
\end{equation}
places the system in symmetry class D \cite{Alt97}. Here $\sigma_x$ is a Pauli matrix that acts on the electron-hole degree of freedom and the complex conjugation operation is taken in the real-space basis (so the momentum $\bm{p}=\hbar\bm{k}=-i\hbar\partial/\partial\bm{r}$ changes sign).

We discretize the Hamiltonian on a 2D square lattice (lattice constant $a$),
\begin{align}
&H=\begin{pmatrix}
\varepsilon_k-\mu&\Delta(\sin ak_x-i\sin ak_y)\\
\Delta(\sin ak_x+i\sin ak_y)&\mu-\varepsilon_k
\end{pmatrix},\nonumber\\
& \varepsilon_k=-t(\cos ak_x+\cos ak_y),\label{Hlattice}
\end{align}
with the definitions $\Delta=(\hbar/a)v_\Delta$, $t=\hbar^2/ma^2$, $\mu=E_{\rm F}-2t$.

We introduce electrostatic disorder by letting the chemical potential $\mu(x,y)$ fluctuate randomly, uniformly distributed in the interval $(\bar{\mu}-\delta\mu,\bar{\mu}+\delta\mu)$. Our approach requires some degree of smoothness of the fluctuating potential on the scale of the lattice constant, in what follows we choose the same $\mu$ on the four neighboring sites $(2n,2m)$, $(2n+1,2m)$, $(2n,2m+1)$, and $(2n+1,2m+1)$.

\subsection{Spectral localizer: open boundary conditions}

The spectral localizer for a two-dimensional class D Hamiltonian with open boundary conditions is \cite{Loz19}
\begin{subequations}
\label{calLdefopen}
\begin{align}
&{\cal L}(x_0,y_0)=\begin{pmatrix}
H&0\\
0&-H
\end{pmatrix}+\kappa \Omega(x-x_0,y-y_0),\\
&\Omega(x,y)=\begin{pmatrix}
0&\sigma_0(x-iy)\\
\sigma_0(x+iy)&0
\end{pmatrix}.
\end{align}
\end{subequations}
The Hermitian operators ${\cal L}$ and $\Omega$ are both $4\times 4$ matrices, we have introduced the $2\times 2$ unit matrix $\sigma_0$ to indicate that $\Omega$ is diagonal in the electron-hole degree of freedom. Note also that $x$ and $y$ are operators (which do not commute with $H$), while $x_0$ and $y_0$ are parameters. Our choice $\kappa=2.5\,t$ for the scale parameter $\kappa$ is explained in App.\ \ref{app_clean}.

The operator $\Omega$ breaks the $\pm E$ symmetry of the spectrum of ${\cal L}$, allowing for a nonzero matrix signature: $\operatorname{Sig}{\cal L}$ = number of positive eigenvalues minus number of negative eigenvalues. This even integer determines a topological invariant, the Chern number \cite{Loz19},
\begin{equation}
{\cal C}(x_0,y_0)=\tfrac{1}{2}\operatorname{Sig}{\cal L}(x_0,y_0),\label{CSigL}
\end{equation}
of a domain containing the point $(x_0,y_0)$. Domain walls, contours across which ${\cal C}(x_0,y_0)$ changes by $\pm 1$, are contours along which $\det{\cal L}(x_0,y_0)$ vanishes. These can be visualized by plotting the \textit{localizer gap} 
\begin{equation}
\delta(x_0,y_0)=\min_n|\lambda_n|,\;\;\lambda_n\;\text{eigenvalue of}\;{\cal L}(x_0,y_0),
\end{equation}
which vanishes along the domain walls.

\subsection{Spectral localizer: periodic boundary conditions}

Our system is a square of size $L\times L$ in the $x$- and $y$-directions. To avoid edge states and focus on bulk properties, we prefer to work with periodic boundary conditions, rather than open boundary conditions. For that purpose, following Ref.\ \onlinecite{Dol24}, the term $x\pm iy$ on the off-diagonal of $\Omega$ is replaced by the periodic combination $\sin(2\pi x/L)\pm i\sin(2\pi y/L)$. The eigenvalues of $\Omega(x-x_0,y-y_0)$ then cannot distinguish between points $x_0$ and $x_0+ L/2$, or between $y_0$ and $y_0+ L/2$. To remove this doubling, cosine terms are added on the diagonal of $\Omega$ \cite{Dol24},
\begin{widetext}
\begin{subequations}
\label{calLdef}
\begin{align}
&{\cal L}(x_0,y_0)=\begin{pmatrix}
H&0\\
0&-H
\end{pmatrix}+\kappa \Omega(x-x_0,y-y_0),\\
&\Omega(x,y)=\begin{pmatrix}
\sigma_0[\cos(2\pi x/L)+\cos(2\pi y/L)-2]&\sigma_0[\sin(2\pi x/L)-i\sin(2\pi y/L)]\\
\sigma_0[\sin(2\pi x/L)+i\sin(2\pi y/L)]&-\sigma_0[\cos(2\pi x/L)+\cos(2\pi y/L)-2]
\end{pmatrix}.
\end{align}
\end{subequations}
\end{widetext}
For $|x|,|y|\ll L$ the localizers \eqref{calLdefopen} and \eqref{calLdef} coincide.

In Fig.\ \ref{fig_landscape} we show the resulting network of domain walls for a particular disorder realization (panels a and b). The topological information contained in the spectral localizer is essential: as shown in panel c, the domain walls do not show up in the local density of states near $E=0$.

\section{Phase diagram from percolation transition}

\subsection{Percolating domain walls}

\begin{figure}[tb]
\centerline{\includegraphics[width=1\linewidth]{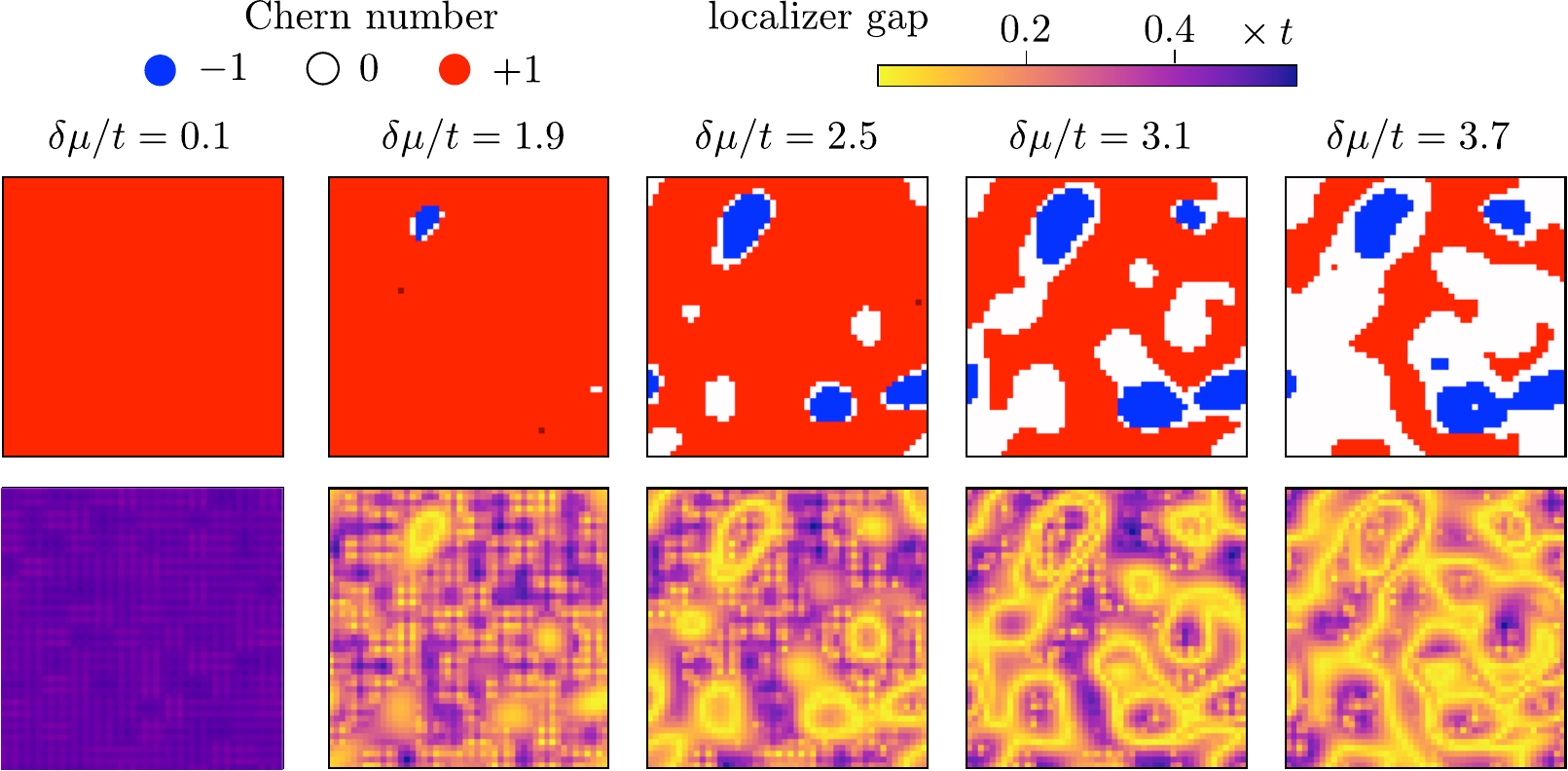}}
\caption{Topological landscape function (Chern number on top row, localizer gap on bottom row) for five different disorder strengths $\delta\mu$ at fixed $\bar{\mu}=1.05\,t$ (and $\Delta=4t$, $L=24a$), to show the appearance of a percolating domain wall when $\delta\mu\gtrsim 3.1\,t$. These are results for a single realization of the random potential $\mu(x,y)$, only the amplitude is rescaled.
}
\label{fig_percolation}
\end{figure}

\begin{figure}[tb]
\centerline{\includegraphics[width=0.5\linewidth]{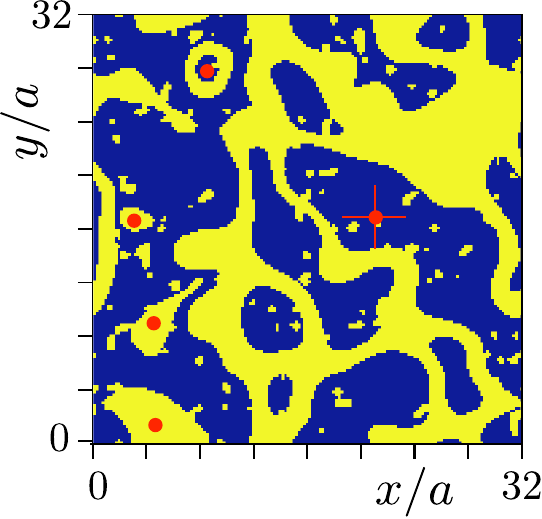}}
\caption{Same as Fig.\ \ref{fig_landscape}b), but now the domain walls are highlighted in yellow, according to the criterion of localizer gap $\delta(\bm{r})<0.1\,t$. The location of the average $\bar{\bm{r}}$ (``center of mass'') for each connected domain wall is indicated by a red dot. The extension $\ell$ of a domain wall is defined by $\ell^2=4\overline{|\bm{r}-\bar{\bm{r}}|^2}$. The red dot marked with a cross identifies the center of mass of a percolating domain wall ($\ell>L$).
}
\label{fig_domainwalls}
\end{figure}

The clean system (without disorder, $\delta\mu=0$) is a topologically trivial thermal insulator (${\cal C}=0$) for $|\bar{\mu}|>2t$ and a topologically nontrivial thermal insulator (${\cal C}=\pm 1$) for $|\bar{\mu}|<2t$. At $\bar{\mu}=0$ there is an insulator-to-insulator transition at which ${\cal C}$ changes sign \cite{Wan21}. Disorder introduces minority domains with a different Chern number than these clean values ${\cal C}_{\rm clean}$. See for example Fig.\ \ref{fig_landscape}, where $\bar{\mu}=t$ and ${\cal C}_{\rm clean}=+1$. 

The domain walls that separate regions of different Chern number support states close to the Fermi level, at energy $E\simeq \hbar v_{\rm F}/\ell$ dictated by the requirement that the kinematic phase upon traveling once around the domain wall cancels the $\pi$ Berry phase. When the extension $\ell$ of the largest domain wall reaches the system size $L$ thermal conduction becomes possible near the Fermi level and the thermal insulator becomes a thermal metal. In Fig.\ \ref{fig_percolation} we show this percolation transition of topological domain walls for a single disorder realization, upon increasing the amplitude $\delta\mu$ of the potential fluctuations at fixed average $\bar{\mu}$.

To identify the percolation transition we need a computationally efficient way to measure the extension $\ell$ of a domain wall. We take localizer gap $\delta(x_0,y_0)<0.1\,t$ as the criterion for a domain wall. All points $\bm{r}=(x_0,y_0)$ satisfying this criterion in a connected region belong to a single domain wall ${\cal D}$. We then compute the domain wall extension $\ell$ from the variance $\sigma^2$ of these points,
\begin{equation}
\ell=2\sigma,\;\;\sigma^2=\overline{|\bm{r}-\bar{\bm{r}}|^2},\label{elldef}
\end{equation}
where $\overline{f(\bm{r})}$ averages a function $f(\bm{r})$ over all $\bm{r}\in{\cal D}$. The procedure is illustrated in Fig.\ \ref{fig_domainwalls}. Our criterion for a \textit{percolating} domain wall is $\ell>L$.

\subsection{Phase diagram}

\begin{figure}[tb]
\centerline{\includegraphics[width=1\linewidth]{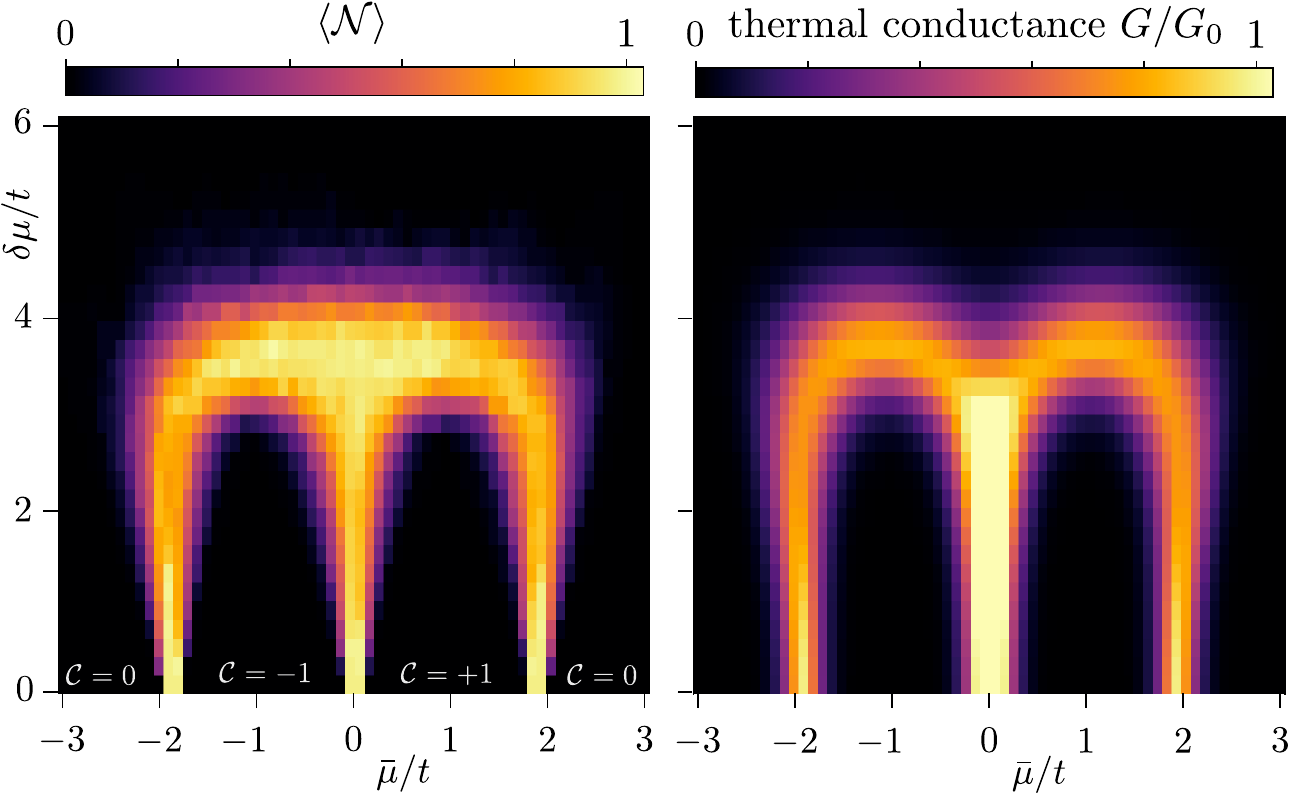}}
\caption{\textit{Left panel:} Color scale plot of the average number $\langle{\cal N}\rangle$ of percolating domain walls, averaged over 100 disorder realizations in the chiral \textit{p}-wave superconductor (parameters $\Delta=4t$, $L=24a$). The value of the Chern number ${\cal C}$ in the clean system ($\delta\mu=0$) is indicated. \textit{Right panel:} Dimensionless thermal conductance for the same system. The uniformly yellow bar at $\bar{\mu}=0$ indicates $G/G_0> 1$.
}
\label{fig_phasediagram}
\end{figure}

The number ${\cal N}$ of percolating domain walls (with $\ell>L$) for a given disorder realization is averaged over the disorder. The resulting dependence of $\langle{\cal N}\rangle$ on the parameters $\bar{\mu}$ and $\delta\mu$ is shown in Fig.\ \ref{fig_phasediagram} (left panel). The region ${\cal N}\approx 1$ where the domain walls percolate is clearly distinguished. 

\begin{figure}[tb]
\centerline{\includegraphics[width=0.7\linewidth]{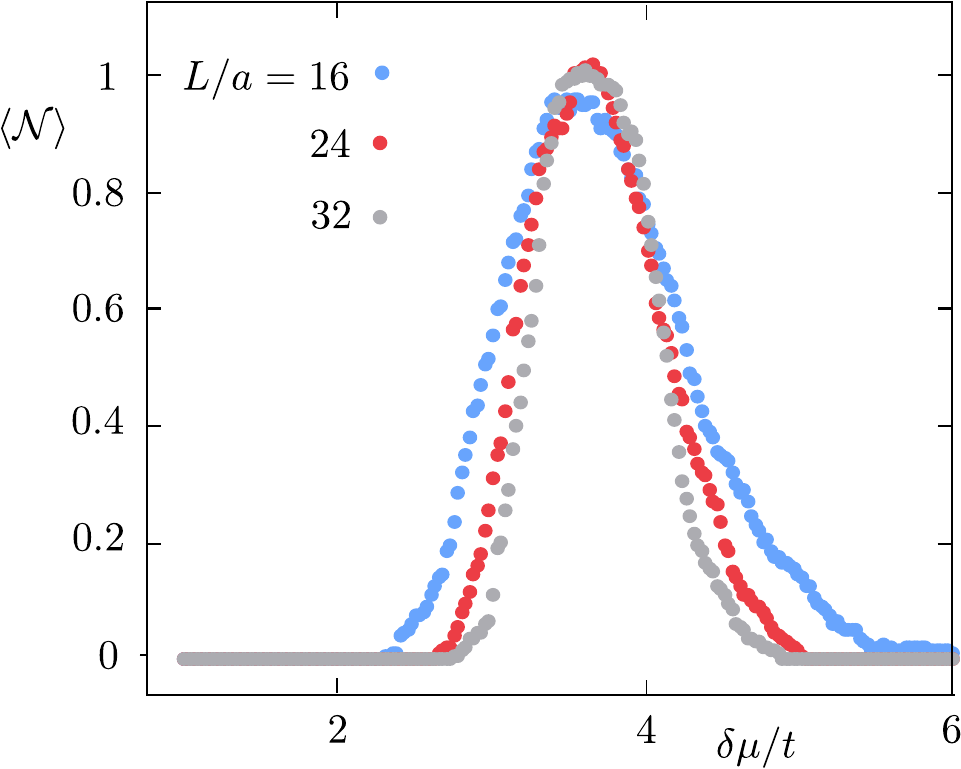}}
\caption{Comparison of the disorder strength dependence of the average number $\langle{\cal N}\rangle$ of percolating domain walls for different system sizes $L$. The data points are averaged over 200 disorder realizations (parameters $\Delta=4t$, $\bar{\mu}=t$). 
}
\label{fig_scaling}
\end{figure}

The data in Fig.\ \ref{fig_phasediagram} is for a relatively small system ($L/a=24$), in Fig.\ \ref{fig_scaling} we compare with a larger system. The critical disorder strength for the percolation transition is approximately scale invariant.

\subsection{Comparison with thermal conductance}

So far we have considered a closed system. If we connect leads at the two ends we can study the thermal conductance,
\begin{equation}
G=G_0\,\operatorname{Tr}\bm{tt}^\dagger,\;\;G_0= \pi^2 k_{\rm B}^2 T/6h,
\end{equation}
at temperature $T$, with $\bm{t}$ the transmission matrix at the Fermi level. The result of such a calculation, using the {\sc kwant} code \cite{kwant}, is also shown in Fig.\ \ref{fig_phasediagram} (right panel).

If we compare with the percolation transition (left panel), we see a good quantitative agreement on the low-disorder side of the phase boundary. The high-disorder side misses a feature in the region near $\bar{\mu}=0$, $\delta\mu=4t$, where the thermal conductance localizes more quickly than inferred from the percolating domain walls. We are unsure about the origin of this difference. Apart from this region the agreement is quite satisfactory, without any adjustable parameters.

\section{Conclusion}

We have shown that the thermal metal phase in a model of a chiral \textit{p}-wave superconductor with electrostatic disorder has a precursor in the thermally insulating phase: The disorder produces domain walls that separate topologically distinct regions (different Chern number). The thermal metal--insulator transition is accompanied by a percolation of the domain walls across the system, providing a transport channel for Majorana fermions (charge-neutral, low-energy excitations).

To reveal the network of domains walls we have used the matrix signature of the spectral localizer \cite{Loz19,Dol24}. We turned to this topological invariant after we were not able to identify localized Majorana fermions using a variation \cite{Lem20,Her20} of the landscape function approach that has been so succesful in the study of Anderson localization \cite{Fil12,Arn16,Fil17,Fil24}. In a sense, the matrix signature of the spectral localizer functions as a \textit{topological} landscape function, sensitive to topological electronic properties that remain hidden in the local density of states.

It would be interesting to study the critical exponent $\nu$ for the percolation transition of the topological domain walls (the exponent that governs the divergence of the largest domain size). Classical 2D percolation has $\nu_{\rm classical}=4/3$. It is suggestive that a recent numerical study \cite{Wan21} of the divergence of the localization length at the thermal metal--insulator transition found $\nu\approx 1.35$, but the proximity to $\nu_{\rm classical}$ may well be accidental.

\acknowledgments

Discussions with the Simons Collaboration on Localization of Waves, in particular with M. Filoche and S. Mayboroda, have motivated us to pursue this project.\smallskip\\ 
C.B. and V.A.Z. received funding from the European Research Council (Advanced Grant 832256).\\
J.T. received funding from the National Science Centre, Poland, within the QuantERA II Programme that has received funding from the European Union's Horizon 2020 research and innovation programme under Grant Agreement Number 101017733, Project Registration Number 2021/03/Y/ST3/00191, acronym {\sc tobits}.\\
I.C.F was supported by the Deutsche Forschungsgemeinschaft (DFG, German Research Foundation) under Germany's Excellence Strategy through the W\"{u}rzburg-Dresden Cluster of Excellence on Complexity and Topology in Quantum Matter--\emph{ct.qmat} (EXC 2147, project-id 390858490)

\subsection*{Data and code availability}
Our computer codes are provided in a Zenodo repository \cite{codes}.

\appendix

\section{Spectral localizer in a clean system}
\label{app_clean}

We have tested the ability of the spectral localizer \eqref{calLdef} to identify the Chern number domains in a clean system, with a smoothly varying $\mu$, where the boundaries are known analytically \cite{Wan21}:
\begin{equation}
{\cal C}=\begin{cases}
0&\text{if}\;\;\mu<-2t,\\
-1&\text{if}\;\;-2t<\mu<0,\\
+1&\text{if}\;\;0<\mu<2t,\\
0&\text{if}\;\;\mu>2t.
\end{cases}
\end{equation}
This test allows us to find a suitable value of the scale parameter $\kappa$. 

\begin{figure}[tb]
	\centerline{\includegraphics[width=0.9\linewidth]{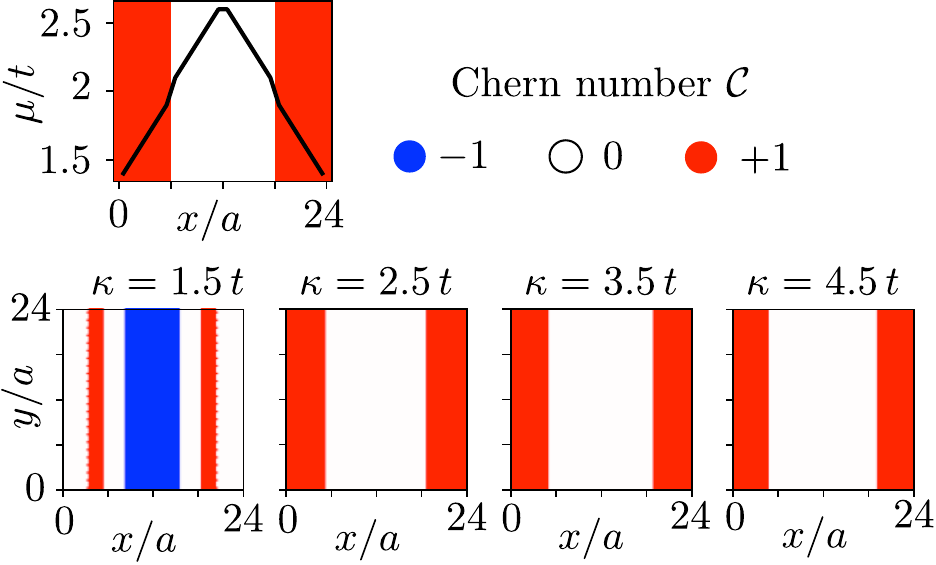}}
	\caption{Top panel: linearly varying $\mu(x)$ (at constant $\Delta=4t$), producing a central domain of Chern number ${\cal C}=0$, flanked by domains of ${\cal C}=+1$. The domain walls are at $x/a=6$ and $x/a=12$. Bottom panels: Chern number domains produced by the spectral localizer, via Eq.\ \eqref{CSigL}, for different values of the scale parameter $\kappa$. We need $\kappa\gtrsim 2t$ for reliable results.
	}
	\label{fig_clean}
\end{figure}

Refs.\ \cite{Loz19,Dol24} argue that $\kappa$ should be of the order of the norm of the Hamiltonian, which in our case is below $10^{-2}\,t$. We find a poor performance for such small $\kappa$, see Fig.\ \ref{fig_clean}, we need $\kappa\gtrsim 2t$ to reliably identify the domain walls. The results in the main text are for $\kappa=2.5\,t$.

\begin{thebibliography}{99}
\bibitem{Kal15} C. Kallin and J. Berlinsky, \textit{Chiral superconductors}, Rep. Prog. Phys. \textbf{79}, 054502 (2016).
\bibitem{Sen00} T. Senthil and M. P. A. Fisher, \textit{Quasiparticle localization in superconductors with spin-orbit scattering}, Phys. Rev. B \textbf{61}, 9690 (2000).
\bibitem{Eve08} F. Evers and A. D. Mirlin, \textit{Anderson transitions}, Rev. Mod. Phys. \textbf{80}, 1355 (2008).
\bibitem{Bee16} C. W. J. Beenakker and L. P. Kouwenhoven, \textit{A road to reality with topological superconductors}, Nature Phys. \textbf{12}, 618 (2016).
\bibitem{Cha02} J. T. Chalker, N. Read, V. Kagalovsky, B. Horovitz, Y. Avishai, and A. W. W. Ludwig, Phys. Rev. B \textbf{65}, 012506 (2001).
\bibitem{Mil07} A. Mildenberger, F. Evers, A.D. Mirlin, and J. T. Chalker, Phys. Rev. B \textbf{75}, 245321 (2007).
\bibitem{Wim10} M. Wimmer, A. R. Akhmerov, M. V. Medvedyeva, J. Tworzyd{\l}o, and C. W. J. Beenakker, \textit{Majorana bound states without vortices in topological superconductors with electrostatic defects}, Phys. Rev. Lett. \textbf{105}, 046803 (2010).
\bibitem{Kag10} V. Kagalovsky and D. Nemirovsky, \textit{Critical fixed points in class D superconductors}, Phys. Rev. B \textbf{81}, 033406 (2010).
\bibitem{Med11} M. V. Medvedyeva, J. Tworzyd{\l}o, and C. W. J. Beenakker, \textit{Effective mass and tricritical point for lattice fermions localized by a random mass}, Phys. Rev. B \textbf{81}, 214203 (2010).
\bibitem{Lau12} C. R. Laumann, A. W. W. Ludwig, D. A. Huse, and S. Trebst, \textit{Disorder-induced Majorana metal in interacting non-Abelian anyon systems}, Phys. Rev. B \textbf{85}, 161301(R) (2012).
\bibitem{Pek19} B. Pekerten, A. Mert Bozkurt, and I. Adagideli, \textit{Fermion parity switches of the ground state of Majorana billiards}, Phys. Rev. B \textbf{100}, 235455 (2019).
\bibitem{Wan21} T. Wang, Z. Pan, T. Ohtsuki, I. A. Gruzberg, and R. Shindou, \textit{Multicriticality of two-dimensional class-D disordered topological superconductors}, Phys. Rev. B \textbf{104}, 184201 (2021).
\bibitem{Car64} C. Caroli, P.-G. de Gennes, and J. Matricon, \textit{Bound fermion states on a vortex line in a type II superconductor}, J. Phys. Lett. \textbf{9}, 307 (1964).
\bibitem{Vol99} G. E. Volovik, \textit{Fermion zero modes on vortices in chiral superconductors}, JETP Lett. \textbf{70}
, 609 (1999).
\bibitem{Rea00} N. Read and D. Green, \textit{Paired states of fermions in two dimensions with breaking of parity and time-reversal symmetries, and the fractional quantum Hall effect}, Phys. Rev. B \textbf{61}, 10267 (2000).
\bibitem{Bee13} C. W. J. Beenakker, \textit{Search for Majorana fermions in superconductors}, Annu. Rev. Con. Mat. Phys. \textbf{4}, 113 (2013).
\bibitem{Lut18} R. M. Lutchyn, E. P. A. M. Bakkers, L. P. Kouwenhoven, P. Krogstrup, C. M. Marcus, and Y. Oreg,\textit{Majorana zero modes in superconductor--semiconductor heterostructures}, Nature Rev. Materials \textbf{3}, 52 (2018).
\bibitem{Fle21} K. Flensberg, F. von Oppen, and A. Stern, \textit{Engineered platforms for topological superconductivity and Majorana zero modes}, Nature Rev. Materials \textbf{6}, 944 (2021).
\bibitem{Lor15} T. A. Loring, \textit{K-theory and pseudospectra for topological insulators}, Annals Physics \textbf{356}, 383 (2015).
\bibitem{Ful16} I. C. Fulga, D. I. Pikulin, and T. A. Loring, \textit{Aperiodic weak topological superconductors}, Phys. Rev. Lett. \textbf{116}, 257002 (2016).
\bibitem{Lor17} T. Loring and H. Schulz-Baldes, \textit{Finite volume calculation of K-theory invariants}, New York J. Math. \textbf{22}, 1111 (2017).
\bibitem{Loz19} E. Lozano Viesca, J. Schober, and H. Schulz-Baldes, \textit{Chern numbers as half-signature of the spectral localizer}, J. Math. Phys. \textbf{60}, 072101 (2019).
\bibitem{Lor20} T. Loring and H. Schulz-Baldes, \textit{The spectral localizer for even index pairings}, J. Noncommutative Geometry \textbf{14}, 1 (2020).
\bibitem{Cer22} A. Cerjan and T. A. Loring, \textit{Local invariants identify topology in metals and gapless systems}, Phys. Rev. B \textbf{106}, 064109 (2022).
\bibitem{Sch24} H. Schulz-Baldes, \textit{Topological indices in condensed matter}, arXiv:2403.18948.
\bibitem{Dol24} N. Doll, T. Loring, and H. Schulz-Baldes, \textit{Local topology for periodic Hamiltonians and fuzzy tori}, arXiv:2403.18931.
\bibitem{Fra24} S. Franca and A. G. Grushin, \textit{Topological zero-modes of the spectral localizer of trivial metals}, Phys. Rev. B \textbf{109}, 195107 (2024).
\bibitem{Qi24} Zihao Qi, Ilyoun Na, Gil Refael, and Yang Peng, \textit{Real-space topological invariant for time-quasiperiodic Majoranas}, Phys. Rev. B \textbf{110}, 014309 (2024).
\bibitem{Vol18} G. E. Volovik, \textit{Topology of a ${}^3$He-A film on a corrugated graphene substrate}, JETP Lett. \textbf{107}, 115 (2018).
\bibitem{Alt97} A. Altland and M. R. Zirnbauer, \textit{Nonstandard symmetry classes in mesoscopic normal-superconducting hybrid structures}, Phys. Rev. B \textbf{55}, 1142 (1997).
\bibitem{kwant} C. W. Groth, M. Wimmer, A. R. Akhmerov, and X. Waintal, \textit{Kwant: A software package for quantum transport}, New J. Phys. \textbf{16}, 063065 (2014).
\bibitem{Lem20} G. Lemut, M. J. Pacholski, O. Ovdat, A. Grabsch, J. Tworzyd{\l}o, and C. W. J. Beenakker, \textit{Localization landscape for Dirac fermions}, Phys. Rev. B \textbf{101}, 081405(R) (2020).
\bibitem{Her20} L. Herviou and J. H. Bardarson, \textit{${\cal L}^2$ localization landscape for highly excited states}, Phys. Rev. B \textbf{101}, 220201(R) (2020).
\bibitem{Fil12} M. Filoche and S. Mayboroda, \textit{Universal mechanism for Anderson and weak localization}, Proc. Natl Acad. Sci. USA \textbf{109}, 14761 (2012).
\bibitem{Arn16} D. N. Arnold, G. David, D. Jerison, S. Mayboroda, and M. Filoche, \textit{Effective confining potential of quantum states in disordered media}, Phys. Rev. Lett. \textbf{116}, 056602 (2016).
\bibitem{Fil17} M. Filoche, M. Piccardo, Y.-R. Wu, C.-K. Li, C. Weisbuch, and S. Mayboroda, \textit{Localization landscape theory of disorder in semiconductors}, Phys. Rev. B \textbf{95}, 144204 (2017).
\bibitem{Fil24} M. Filoche, P. Pelletier, D. Delande, and S. Mayboroda, \textit{Anderson mobility edge as a percolation transition}, Phys. Rev. B \textbf{109}, L220202 (2024).
\bibitem{codes} The simulation code for the spectral localizer calculations is available in the Zenodo repository at \url{https://doi.org/10.5281/zenodo.14782907}.
\end{thebibliography}
\end{document}